\documentclass[twocolumn,preprintnumbers,amsmath,amssymb]{revtex4}

\usepackage{graphicx}
\usepackage{dcolumn}
\usepackage{bm}

\def\bea{\begin{eqnarray}}
\def\eea{\end{eqnarray}}

\begin{document} 



\title{Comment on ``Exposing the non-collectivity in elliptic flow''}

\author{Thomas A. Trainor}
\address{CENPA 354290, University of Washington, Seattle, WA 98195}


\date{\today}

\begin{abstract}
\end{abstract}


\maketitle


A method is proposed to distinguish ``elliptic flow'' from non-collective (jet) contributions using so-called ``forward-backward'' (FB) correlations on pseudorapidity $\eta$~\cite{koch}. The proposed method has two issues: 1) The FB technique approximates 2D angular autocorrelations~\cite{autocorr} having extensive applications (e.g.~\cite{axialci,axialcd,ptscale,daugherity}). 2) Assumptions about jet systematics invoked to support the method fail dramatically in more-central RHIC Au-Au collisions where separation of  nonjet quadrupole (``elliptic flow'') and jet angular correlations is most critical. Accurate separation of the nonjet quadrupole from jets is achieved by fits to 2D angular autocorrelations~\cite{gluequad,quadspec,kettler}. Measured jet systematics reveal strong elongation on $\eta$ of the same-side jet peak~\cite{axialci,daugherity}.


The proposed FB method estimates correlations of quadrupole amplitudes in two $\eta$ bins (F and B) symmetrically placed about the collision center of momentum (CM). The main assumptions of the method are: 1) At most one jet appears in each collision within a single $\eta$ bin, whereas the collective part (nonjet quadrupole, ``elliptic flow'') is common to both. 2) The ``away-side'' jet is absent (is ``quenched''). 3) Jets are narrow on (pseudo)rapidity, do not contribute to both F and B bins. 


Eq.~(4) of~\cite{koch} can be rewritten in terms of particle multiplicities $N$ in bins centered at $(p_{t},\eta,\phi)$ with $dN(p_{t},\eta,\phi)/d\phi \sim N(p_{t},\eta,\phi)/ \delta \phi$ (bin width $\delta \phi$) as
\bea
2\left\langle V_2(p_{t1},\eta_1)\, V_2(p_{t2},\eta_2) \right\rangle \hspace{-.05in}  &=& \hspace{-.05in}  \int \hspace{-.07in}  \int_0^{2\pi} \hspace{-.1in}  d\phi_1 d\phi_2\,  \cos(2[\phi_1 - \phi_2]) \nonumber \\
&&\hspace{-1.4in} \times  \left \langle [dN(p_{t1},\eta_1,\phi_1)/d\phi_1] \,
 [dN(p_{t2},\eta_2,\phi_2)/d\phi_2] \right \rangle,
\eea
where ``FB'' implies $\eta_1 = - \eta_2$. Other $(\eta_1,\eta_2)$ combinations are possible.  In case {\em stationarity} on $\eta$ holds (invariance on $\eta_\Sigma = \eta_1 + \eta_2$, a good approximation within the STAR TPC acceptance~\cite{axialcd}) the ensemble mean (brackets) on the RHS is an element of a 2D angular autocorrelation on $(\eta_\Delta,\phi_\Delta)$, with e.g. $\phi_1 - \phi_2 \equiv \phi_\Delta$. The azimuth integral is the quadrupole or $m=2$ Fourier amplitude of the total 2D angular autocorrelation for a particular combination $(\eta_1,\eta_2) \rightarrow \eta_\Delta$. The FB configuration assumed for the proposed method implies $\eta_\Sigma = 0$, which is not a necessary condition and unnecessarily reduces the available statistical power within the detector acceptance. Averaging over $\eta_\Sigma$ generates the full autocorrelation~\cite{autocorr}.

To obtain $\left\langle V_2^F\, V_2^B \right \rangle(p_t)$ in a manner consistent with conventional $v_2(p_t)$ analysis an integral of $\left\langle V_2(p_{t1},\eta_1)\, V_2(p_{t2},\eta_2 \right \rangle$ over $p_{t1}$ (and symmetrically over $p_{t2}$) with $\eta_2=-\eta_1$ must be performed to obtain a {\em marginal} distribution on $p_t$~\cite{quadspec}. A similar procedure with $\eta_1 = \eta_2$ produces $\langle [V_2^{F}]^2\rangle(p_t) = \langle [V_{2}^{B}]^2\rangle(p_t)$, assuming collision symmetry about the CM. Correlation coefficient $C_{FB}(p_t) \equiv \left\langle V_2^F\, V_2^B \right \rangle(p_t) / \sqrt{\langle [V_2^F]^2 \rangle(p_t) \, \langle [V_2^B]^2\rangle(p_t) }$ measures the FB correlation of the azimuth quadrupole amplitude, with 1 expected for a true collective phenomenon and 0 expected for resolved single jets.

Azimuth correlations are decomposed into nonjet quadrupole component $\cal F$ and jet contribution ${\cal J}$. Parameter $g(p_t)$ is the fractional jet quadrupole amplitude.  $v_2(p_t)$ is then expressed as $v_2(p_t) = (1-g)\, v_2^{\cal F} + g\, v_2^{\cal J}$ denoting nonjet (``flow'') and jet (``nonflow'') contributions. The expression for $C_{FB}(p_t)$ in Eq.~(10) of~\cite{koch} would be 1 except for the missing jet-jet term in the numerator, which is set to zero based on the assumption of at most one jet in each collision, with $\eta$ width less than the F-B bin spacing $|2\eta_1|$. That assumption fails dramatically for more-central Au-Au collisions. There are many jets~\cite{fragevo}, and same-side jet peaks extend across several units of $\eta$, are thus common to F and B $\eta$ bins~\cite{daugherity}. Consequently, $C_{FB}(p_t) \sim 1$ from jets in more-central Au-Au collisions would seem to confirm a large degree of  ``collectivity''.

The nonjet quadrupole can be separated accurately from jet correlations by  model fits to 2D angular autocorrelations~\cite{gluequad,quadspec,kettler}. 
The jet-related quadrupole~\cite{zyam} extends down to small $p_t$ values, consistent with quantitative pQCD descriptions of jet structure on $p_t$~\cite{fragevo}.  The $p_t$ dependence of the nonjet quadrupole component of minimum-bias $v_2(p_t)$ data is well-described by a L\'evy {\em quadrupole spectrum} shape boosted by $\langle \beta_t \rangle \sim 0.6$~\cite{quadspec}.




\begin{thebibliography}{99}

\bibitem{koch} J.~Liao and V.~Koch,
Phys.~Rev.~Lett.~{\bf 103}, 042302 (2009).

\bibitem{autocorr} T.\,A.~Trainor, R.\,J.~Porter and D.\,J.~Prindle, J. Phys. G: Nucl. Part. Phys.  {\bf 31}, 809 (2005). 

\bibitem{axialci} J.~Adams {\it et al.}  (STAR Collaboration),
  Phys.\ Rev.\  C {\bf 73}, 064907 (2006).

\bibitem{axialcd}  J.~Adams {\it et al.}  (STAR Collaboration),
  Phys.\ Lett.\  B {\bf 634}, 347 (2006).

\bibitem{ptscale} J. Adams {\it et al.} (STAR Collaboration), J.~Phys.~G.: Nucl. Part. Phys. {\bf 32}, L37 (2006). 

\bibitem{daugherity}  M.~Daugherity  (STAR Collaboration),
  J.\ Phys.\ G {\bf 35}, 104090 (2008).

\bibitem{gluequad}  T.~A.~Trainor,
  Mod.\ Phys.\ Lett.\  A {\bf 23}, 569 (2008).

\bibitem{quadspec}    T.~A.~Trainor,
  Phys.\ Rev.\  C {\bf 78}, 064908 (2008).

\bibitem{kettler}  D.~T.~Kettler  (STAR collaboration),
  Eur.\ Phys.\ J.\  C {\bf 62}, 175 (2009),
  arXiv:0907.2686.

\bibitem{fragevo}  T.~A.~Trainor,
  arXiv:0901.3387.

\bibitem{zyam}  T.~A.~Trainor,
  arXiv:0904.1733.

\end{thebibliography}
\end{document}